\documentstyle[12pt,epsfig]{article}
\begin{document}
\newcommand{\beq}{\begin{equation}}
\newcommand{\eeq}{\end{equation}}
\newcommand{\bea}{\begin{eqnarray}}
\newcommand{\eea}{\end{eqnarray}}
\newcommand{\lam}{\lambda}
\newcommand{\GE}{\gamma_{e}}
\newcommand{\gsim}{\raisebox{-0.07cm}{$\:\stackrel{>}{{\scriptstyle
 \sim}}\: $} }
\newcommand{\lsim}{\raisebox{-0.07cm}{$\:\stackrel{<}{{\scriptstyle
 \sim}}\: $} }
\setlength{\baselineskip}{0.58cm}
\setlength{\parskip}{0.4cm}
\begin{titlepage}

\noindent
{\tt hep-ph/9910545} \hfill INLO-PUB 19/99 \\
\hspace*{\fill} October 1999 \\
\vspace{1.5cm}
\begin{center}
\Large
{\bf On Soft Gluon Effects in} \\
\vspace{0.1cm}
{\bf Deep-Inelastic Structure Functions} \\
\large
\vspace{2.8cm}
A. Vogt \\
\vspace{1.2cm}
\normalsize
{\it Instituut-Lorentz, University of Leiden \\
\vspace{0.1cm}
 P.O. Box 9506, 2300 RA Leiden, The Netherlands} \\
\vfill
\large
{\bf Abstract} \\
\end{center}
\vspace{-0.3cm}
\normalsize
The behaviour of the quark coefficient functions in deep-inelastic
scattering is investigated for large values of the Bjorken variable 
$x$. By combining results of soft-gluon resummation and fixed-order
calculations, we determine the coefficients of the four leading
large-$x$ logarithms, $\alpha_s^k\, [\{ \ln (1-x)\} ^{2k-l}/(1-x)]_+$, 
$l = 1, \ldots 4$, to all orders in the strong coupling constant
$\alpha_s$. This result includes two more terms for the three-loop 
coefficient functions than previously specified in the literature. 
The effect of the fifth logarithmic contribution is approximately 
evaluated. The terms derived here are required, but also seem to be 
sufficient, for a reliable representation of the coefficient functions 
at large $x$. 

\vspace*{0.5cm}
\normalsize

\end{titlepage}

\noindent
Knowledge of the proton's quark distributions at large momentum 
fractions $x$ is relevant to the search for new phenomena at 
high-energy colliders like HERA, {\sc Tevatron} and the future LHC. 
These distributions are usually inferred from data on structure 
functions in deep-inelastic scattering (DIS). For instance, data even 
up to $x=0.98$ have recently been employed \cite{BY98} to scrutinize a 
possible non-standard behaviour of the large-$x$ quark distributions 
\cite{qmod} suggested in connection with the initial HERA high-$Q^2$ 
anomaly \cite{HERA}.
In perturbative QCD the link between the quark distributions and the 
structure functions $F_i(x,Q^2)$, $i = 1, 2, 3$, is given by the 
coefficient functions$\, $\footnote
{The renormalization and factorization scales are chosen as 
 $\mu_r^2 = \mu_f^2 = Q^2$ throughout this paper. The additional terms 
 arising for other choices can be deduced from renormalization-group 
 constraints.}
\beq
\label{coeffx}
  C_{i,q}(x,Q^2) \: =\:\delta (1-x) + \sum_{k=1} a_s^k c_{i,q}^{(k)}(x)
\eeq
with $a_s = \alpha_s(Q^2)/(4 \pi)$. The expansion coefficients 
$c_{i,q}^{(k)}(x)$ and their Mellin-$N$ space counterparts
$c_{i,q}^{(k)N}$ contain universal (i.e., $i$-independent) soft-gluon 
contributions of the~form
\beq
\label{logs}
  \left[ \frac{\ln^{l-1} (1-x)}{1-x} \right]_+ 
  \:\:\: \longleftrightarrow \:\:\: \frac{(-1)^l}{l}\,
  ( \ln ^l N + \mbox{ subleading terms} ) \:\: ,
\eeq
$l = 1, \ldots , 2k$. At sufficiently large $x\, $/$\, $large $N$, 
these terms spoil the convergence of finite-order approximations to 
$C_{i,q}$ and need to be resummed to all orders. It is thus important 
to derive a reliable estimate of their impact at the $x$-values 
included in data analyses.

The soft-gluon resummation has been carried out for the leading and 
next-to-leading logarithms (in the sense of eq.~(\ref{res2}) below) 
some time ago \cite{NLL1,NLL2}. In this letter we combine these 
results with terms from the one- and two-loop coefficient functions 
$c_{i,q}^{(1,2)}(x)$ calculated in refs.~\cite{cq1,cq2}. This 
information fixes the first four towers, $j = 0 ,1, 2$ and 3, of 
large-$x$ logarithms, i.e., the coefficients of $\ln^{2k-j} N$ in 
$c_{i,q}^{(k)N} $at each order $k$. It also facilitates an estimate of 
the effect of the fifth tower, $j=4$. As shown below, these four to 
five towers are required, but also seem to be sufficient, for a 
realistic estimate of the higher-order effects at large $x$.

Up to terms which vanish for $N \rightarrow \infty$, the $N$-space 
coefficient functions take the form  
\beq
\label{coeffN}
  C_{i,q}^N(Q^2) \: =\: (1 + a_s g_{01}^{} + a_s^2 g_{02}^{} + \ldots ) 
                        \cdot \exp \, [G^N(Q^2)] \:\: .
\eeq
The first factor collects contributions which are constant for 
$N \rightarrow \infty$. Adopting the $\overline{\mbox{MS}}$ scheme, the 
first-order term reads~\cite{cq1}
\beq
\label{g01n}
 g_{01}^{} \: = \: ( - 9 - 2\,\zeta_2^{} + 2\, \GE^2 + 3\, \GE )\, C_F
 \:\: .
\eeq
Here $\zeta_l^{}$ stands for Riemann's $\zeta$-function, $\gamma_e$ is
the Euler-Mascheroni constant, and $C_F^{} = 4/3$ in QCD. The  
corresponding second-order term can be readily inferred from 
ref.~\cite{cq2}.
The function $G^N(Q^2)$ is the object of the soft-gluon resummation. 
It is given by \cite{NLL1,NLL2}
\bea
\label{res1}
  G^N(Q^2) &\! =\! & \int_0^1 \! dz \,\frac{z^{N-1}-1}{1-z} \,\left[
  \int_{Q^2}^{(1-z)Q^2} \frac{dq^2}{q^2}\, A(a_s(q^2)) + \frac{1}{2}\, 
  B(a_s([1-z]Q^2)) \right] 
  \nonumber \\[1mm]
  & & \mbox{}+ {\cal O}(a_s(a_s \ln N)^k) 
\eea
with $A(a_s) = 4\, C_F\, a_s + 8\, C_F K \, a_s ^2 + \ldots $, 
$B = -6\, C_F\, a_s + \ldots $ and
\beq
  K \: =\: \left( \frac{67}{18} - \zeta_2^{} \right) C_A^{} 
      - \frac{5}{9}\, N_f^{} \:\: .
\eeq
$C_A = 3$ in QCD, and $N_f$ denotes the number of effectively massless
quark flavours. 
These terms of $A(a_s)$ and $B(a_s)$, together with the lowest two coefficients of the $\beta$-function, 
\beq
\label{beta}
  \beta_0^{} \: =\: \frac{11}{3}\, C_A^{} - \frac{2}{3}\, N_f^{} 
  \:\: , \quad\quad 
  \beta_1^{} \: =\: \frac{34}{3}\, C_A^2 - \frac{10}{3}\, C_A^{}N_f^{}
                    - 2\, C_F^{}N_f^{} 
\eeq
are sufficient for deriving the next-to-leading logarithms. 

The integrals occurring in eq.~(\ref{res1}) can be found in 
ref.~\cite{CMN}.
Using the abbreviations $L \equiv \ln N$ and $\lam \equiv \beta_0 
a_s L$, the result can be written as
\beq
\label{res2}
  G^N(Q^2) = L\, g_1^{}(a_sL) + g_2^{}(a_sL) + a_s\, g_3^{}(a_sL) 
  + \ldots 
\eeq
with
\bea
\label{g1n}
  g_1^{}(a_sL) &\! =\! & \frac{4 C_F}{\beta_0 \lam}\,
    \Big[ \lam + (1-\lam) \ln(1-\lam) \Big] 
    \:\:\equiv\:\: \sum_{k=1} g_{1k}^{}(a_sL)^k
    \nonumber \\
    &\! =\! & \sum_{k=1}^{\infty} \frac{4\, C_F \beta_0^{k-1}}{k(k+1)} 
    \, (a_sL)^k \:\: , \quad \\[1.5mm]
\label{g2n}
  g_2^{}(a_sL) &\! =\! & - \frac{C_F (3+4\GE)}{\beta_0}\, \ln(1-\lam)
    - \frac{8C_F K}{\beta_0^2}\, \Big[ \lam + \ln(1-\lam)\Big]
  \nonumber \\[1mm]
  & & \mbox{}+ \frac{4 C_F \beta_1}{\beta_0^3}\,
    \Big[ \lam + \ln (1-\lam) + \frac{1}{2}\ln^2(1-\lam) \Big]  
    \:\:\equiv\:\: \sum_{k=1} g_{2k}^{}(a_sL)^k 
  \\[1mm]
  &\! =\! & \sum_{k=1}^{\infty} \left\{ \frac{3+4\GE}{\beta_0}
    + \theta_{k2} \left( \frac{8 K}{\beta_0^{2}} 
    + \frac{4 \beta_1^{}}{\beta_0^{3}} \Big[ S_1(k-1)-1 \Big] 
      \right) \right\} \frac{C_F \beta_0^k}{k} \, (a_s L)^k \:\: .
  \nonumber
\eea
Here $\theta_{kj} = 1$ for $k\geq j$ and $\theta_{kj} = 0$ else, and 
$S_1(k) = \sum_{j=1}^{k} 1/j$. 
The function $g_3^{}$ in eq.~(\ref{res2}),
\beq
  g_3^{}(a_sL) \: \equiv \: \sum_{k=2} g_{3k}^{}(a_sL)^{k-1} \:\: ,
\eeq
is presently unknown except 
for its leading term which can be determined from ref.~\cite{cq2}. 
Its coefficient $g_{32}^{}$ reads
\bea
\label{g32n}
  g_{32}^{} &\! =\! & 
    \left( \frac{3155}{54} - 40\, \zeta_3^{} - \frac{22}{3}\, 
    \zeta_2^{} - 8\,\zeta_2^{} \GE + \frac{22}{3}\, \GE^2 
    + \frac{367}{9}\, \GE \right) C_F^{} C_A^{} 
  \nonumber \\[1mm] & & \mbox{}
  \left( \frac{3}{2} + 24\, \zeta_3^{} - 12\, \zeta_2^{} \right) C_F^2 
    + \left( -\frac{247}{27} + \frac{4}{3}\, \zeta_2^{} - \frac{4}{3}\, 
    \GE^2 - \frac{58}{9}\, \GE \right) C_F^{} N_f^{} \:\: .
\eea

We are now ready to evaluate the resummed large-$N$ coefficient 
function (\ref{coeffN}). Expansion of the exponential yields
\beq
\label{coeffE}
  C_{i,q}^N(Q^2) \: =\:  1 + \sum_{k=1}^{\infty} a_s^k \, 
  \Big( c_{k1}^{} L^{2k} + c_{k2}^{} L^{2k-1} + c_{k3}^{} L^{2k-2} 
  + c_{k4}^{} L^{2k-3} + {\cal O}(L^{2k-4}) \Big) 
\eeq
with
\bea
\label{towers}
  c_{k1}^{} &\! =\! & \frac{g_{11}^k}{k!} 
    \nonumber \\
  c_{k2}^{} &\! =\! & \frac{g_{11}^{k-1}}{(k-1)!}\, g_{21}^{}
    + \frac{\theta_{k2}\, g_{11}^{k-2}}{(k-2)!}\, g_{12}^{} 
    \nonumber \\
  c_{k3}^{} &\! =\! & \frac{g_{11}^{k-1}}{(k-1)!}\, g_{01}^{}
    + \frac{\theta_{k2}\, g_{11}^{k-2}}{(k-2)!}\, \Big( g_{22}^{} 
      + \frac{1}{2} g_{21}^2 \Big) 
    + \frac{\theta_{k3}\, g_{11}^{k-3}}{(k-3)!}\, \Big( g_{13}^{} 
      + g_{12}^{} g_{21}^{} \Big) 
    + \frac{\theta_{k4}\, g_{11}^{k-4}}{2(k-4)!}\, g_{12}^2 
    \nonumber \\
  c_{k4}^{} &\! =\! & \frac{\theta_{k2}\, g_{11}^{k-2}}{(k-2)!}\, 
      \Big( g_{21}^{} g_{01}^{} + g_{32} \Big)
    + \frac{\theta_{k3}\, g_{11}^{k-3}}{(k-3)!}\, \Big( g_{12}^{} 
      g_{01}^{} + g_{23}^{} + g_{22}^{} g_{21}^{} 
      + \frac{1}{6} g_{21}^3 \Big) 
    \nonumber \\ & & \mbox{}
    + \frac{\theta_{k4}\, g_{11}^{k-4}}{(k-4)!}\, \Big( g_{14}^{} 
      + g_{13}^{} g_{21}^{} + g_{12}^{} g_{22}^{} + \frac{1}{2} 
      g_{12}^{} g_{21}^2 \Big) 
    + \frac{\theta_{k5}\, g_{11}^{k-5}}{(k-5)!}\, \Big( g_{13}^{} 
      g_{12}^{} + \frac{1}{2} g_{12}^2 g_{21}^{} \Big)
    \nonumber \\ & & \mbox{}
    + \frac{\theta_{k6}\, g_{11}^{k-6}}{6(k-6)!}\, g_{12}^3  \:\: .
\eea 
Inspection of eq.~(\ref{towers}) reveals that the coefficients up to 
$c_{k4}^{}$ are indeed fixed by eqs.~(\ref{g01n}), (\ref{g1n}), 
(\ref{g2n}) and (\ref{g32n}).\footnote
{The agreement of the predictions (\ref{towers}) for $c_{21}^{}$, 
 $c_{22}^{}$ and $c_{23}^{}$ with the results of the full ${\cal O}
 (a_s^2)$ calculation~\cite{cq2} provides a non-trivial check of the 
 resummation (\ref{res2})--(\ref{g2n}) of the next-to-leading 
 logarithms.}
Especially the four leading large-$x$ terms of the three-loop ($k=3$) 
coefficient functions $c_{i,q}^{(3)}$ are thus determined. 
Transformation to $x$-space \cite{BK98} yields
\bea 
\label{cq3x}
  c_{i,q}^{(3)}(x) &\! =\! & 
    8\, C_F^3 \left[ \frac{\ln^5 (1-x)}{1-x} \right]_+
    - \left( 30\, C_F^3 + \frac{220}{9}\, C_F^2 C_A^{} - \frac{40}{9}\, 
       C_F^2 N_f^{} \right) \left[ \frac{\ln^4 (1-x)}{1-x} \right]_+ 
    \nonumber \\[1mm] & & \mbox{} 
    + \left\{ \bigg( - 36 - 96\, \zeta_2^{} \bigg) \, C_F^3
    + \bigg( \frac{1732}{9} - 32\, \zeta_2^{} \bigg) \, C_F^2 C_A^{}
    + \frac{484}{27}\, C_F^{} C_A^2 
    \right. \nonumber \\ & & \left. \mbox{} \quad\quad
    + \frac{16}{27}\, C_F^{} N_f^2
    - \frac{176}{27}\, C_F^{} C_A^{} N_f^{} 
    - \frac{280}{9}\, C_F^2 N_f^{} \right\} 
    \left[ \frac{\ln^3 (1-x)}{1-x} \right]_+ \, 
    \nonumber \\[1mm] & & \mbox{} 
    + \left\{ \bigg( \frac{279}{2} + 16\, \zeta_3^{} + 288\, \zeta_2^{} 
      \bigg) C_F^3
    + \bigg( -\frac{8425}{18} + 240\, \zeta_3^{} + \frac{724}{3} 
      \zeta_2^{} \bigg) C_F^2 C_A^{} 
    \right. \nonumber \\ & & \left. \mbox{} \quad\quad
    + \bigg( -\frac{4649}{27} + \frac{88}{3}\, \zeta_2^{} \bigg) 
       C_F^{} C_A^2
    + \bigg( \frac{683}{9} - \frac{112}{3}\, \zeta_2^{} \bigg) 
       C_F^2 N_f^{}
    \right. \nonumber \\ & & \left. \mbox{} \quad\quad
    + \bigg( \frac{1552}{27} - \frac{16}{3}\, \zeta_2^{} \bigg) 
       C_F^{} C_A^{} N_f^{}
    - \frac{116}{27}\, C_F^{} N_f^2 \right\} 
    \left[ \frac{\ln^2 (1-x)}{1-x} \right]_+ + \ldots \:\: . \quad
\eea
The first two terms of eq.~(\ref{cq3x}) have already been given in 
ref.~\cite{LRVN}. Besides as a useful cross check for a future exact
calculation of $c_{i,q}^{(3)}(x)$, the above result can also be used 
for improved approximate reconstructions \cite{NV2} of this function 
along the lines of \mbox{refs.~\cite{LRVN,NV1}}. 

\noindent
The corresponding higher-order results are straightforward if 
cumbersome. The numerical values of the coefficients $c_{kl}^{}$ in 
eq.~(\ref{towers}) are presented in table 1 for $N_f = 4$ and 
$k\leq 10$. In contrast to $g_2$ in eq.~(\ref{g2n}) which exhibits a 
pole at $N = \exp [1/(\beta_0 a_s)]$, the sum (\ref{coeffE}) converges 
for all $N$ as long as the number of towers included is finite.

\begin{table}[htb]
\begin{center}
\begin{tabular}{|r||r|r|r|r|r|}\hline
     &           &           &           &           &          \\[-4mm]
 $k$ & $c_{k1}^{} \quad $ & $c_{k2}^{} \quad $ & 
       $c_{k3}^{} \quad $ & $c_{k4}^{} \quad $ & 
       $c_{k5}^{} \quad\quad\quad $                             \\[1mm]
                                                         \hline \hline
     &           &           &           &           &          \\[-4mm]
  1  &  2.66667  &   7.0785  &---$\quad $&---$\quad $& 
                                ---$\quad\quad\quad $           \\[1mm]
  2  &  3.55556  &  26.2834  &   40.760  & $-$67.13  & 
                                ---$\quad\quad\quad $           \\[1mm]
  3  &  3.16049  &  44.9210  &  238.885  &   470.82  &  
                                $ g_{33}^{} -1235.23$           \\[1mm]
  4  &  2.10700  &  48.7090  &  477.854  &  2429.46  &  
                                $ c_{11}\, g_{33}^{} + 3600.12$ \\[1mm]
  5  &  1.12373  &  38.3254  &  581.518  &  5015.18  & 
                                $ c_{21}\, g_{33}^{} + 22963.9$ \\[1mm]
  6  &  0.49944  &  23.5617  &  505.972  &  6432.95  & 
                                $ c_{31}\, g_{33}^{} + 50185.0$ \\[1mm]
  7  &  0.19026  &  11.8592  &  340.954  &  5933.61  & 
                                $ c_{41}\, g_{33}^{} + 67307.5$ \\[1mm]
  8  &  0.06342  &   5.0464  &  186.822  &  4244.45  & 
                                $ c_{51}\, g_{33}^{} + 64858.9$ \\[1mm]
  9  &  0.01879  &   1.8583  &   86.041  &  2467.72  & 
                                $ c_{61}\, g_{33}^{} + 48498.6$ \\[1mm]
 10  &  0.00501  &   0.6028  &   34.118  &  1204.34  & 
                                $ c_{71}\, g_{33}^{} + 29487.7$ \\[1mm]
                                                       \hline
\end{tabular}
\vspace{1mm}
\caption{Numerical values of the coefficients $c_{kl}^{}$ in
 eq.~(\ref{towers}) for $N_f = 4$. Also shown are the corresponding
 (incomplete) results $c_{k5}^{}$ for the fifth tower of logarithms in 
 eq.~(\ref{coeffE}).} 
\end{center}
\end{table}
 
Before turning to the effect of these higher-order contributions, it is 
instructive to compare the $\alpha_s^2$ and $\alpha_s^3$ parts of 
eq.~(\ref{coeffE}) with the respective exact and approximate results 
for $c_{2,q}^{(2)}$ \cite{cq2} and $c_{2,q}^{(3)}$ \cite{LRVN,talk}.%
\footnote
{These approximations are based on the first five even-integer moments 
 of $c_{2,q}^{(3)}(x)$ \cite{LRVN}, supplemented by the first two
 coefficients in eq.~(\ref{cq3x}). $c_{2,q}^{(3)}(x)$ is rather tightly
 constrained at $x \gsim 0.5$ by this information.} 
For this purpose the coefficient function is convoluted (for $\alpha_S
= 0.2$, a value typical for scales probed in fixed-target DIS) with a 
simple, but characteristic model $f(x)$ of the large-$x$ quark 
distributions. The second-order comparison is shown in fig.~1, its 
third-order counterpart in fig.~2. Two and three terms in the large-$N$ 
expansion (\ref{coeffE}) (i.e., $\ln^4 N$ and $\ln^3 N$ at two-loop 
and $\ln^6 N$, $\ln^5 N$, and $\ln^4 N$ at three-loop) turn out to be 
sufficient for good approximations to $c_{2,q}^{(3)} \otimes f$
and $c_{3,q}^{(3)} \otimes f$, respectively, at large $x$. It 
should be noted that it is essential for this fast convergence to use
the expansion in $N$-space. For instance, $c_{2,q}^{(2)} \otimes f$
is severely overestimated if only the two leading $x$-space 
terms, $[ \ln^3 (1-x)/(1-x)]_+$ and $[ \ln^2 (1-x)/(1-x)]_+$, are kept 
\cite{LRVN}. In the all-order case, the problem of the $x$-space 
expansion has been elucidated in ref.~\cite{CMNT}.

The higher-order soft-gluon corrections, $k\geq 4$ in eq.\ 
(\ref{coeffE}), are illustrated in the same manner in fig.~3. Under 
these conditions the terms up to $k = 8$ are sufficient for an 
accurate representation up to $z = 0.99$. As above the Mellin 
inversions are performed using the standard contour~\cite{cont} (see 
also the discussion of the infinite-order limit in ref.~\cite{CMNT})
\bea
  A(x) &\: =\: & \frac{1}{2\pi i} \left\{ \int_{c-(i+\delta)\infty}^{c}
    + \int_{c}^{c+(i-\delta)\infty} \right\} dN\, x^{-N} A_N 
  \nonumber \\[1mm]
  &\: =\: &   \frac{1}{\pi} \int_0^\infty \! dz \: {\rm Im}\, \Big[ 
    \exp (i\phi)\, x^{-c -z \exp (i\phi)}\, A_{N = c +z \exp (i\phi)} 
    \Big]
\eea
using $c = 2$ and $\phi = 3/4\, \pi$ corresponding to $\delta = 1$. The
fourth tower of logarithms, $\sum_{k} c_{k4\,}^{} a_s^k\ln^{2k-3}\! N$, 
yields a very large contribution, e.g., it exceeds the effect of the 
first three towers by a factor of 1.5 at $x = 0.9$. In order to assess 
the convergence of the large-$N$ expansion, it is useful to notice that 
the fifth tower, $\sum_{k} c_{k5\, }^{} a_s^k \ln^{2k-4}\! N$, is 
fixed by available information (including the parameter $g_{02}^{}$ in 
eq.~(\ref{coeffN})), except for the second term $g_{33}^{} 
(a_s \ln N)^2$ of the function $g_3^{}$ in eq.~(\ref{res2}). As shown 
in table 1, the impact of the unknown constant $g_{33}^{}$ (of which 
the order of magnitude can be inferred from $k=3$) is small at 
$k\geq 5$. The effect of the fifth tower can thus be estimated, and 
turns out to be rather moderate, e.g., about 25\% at $x = 0.9$. This
indicates that four to five towers are sufficient for a valid estimate
of the soft-gluon corrections to the quark coefficient functions at 
large~$x$.

To summarize, we have determined the four leading towers of 
large-$x\, $/$\, $large-$N$ soft-gluon logarithms entering the quark 
coefficient functions in DIS, and estimated the fifth tower. There
is evidence that -- provided the resummation is applied in $N$-space,
avoiding uncontrolled lower $\ln^l N$ terms due to the
Mellin transform (\ref{logs}) -- these four to five towers are 
sufficient for a reliable estimate of the higher-order contributions at 
large $x$. These terms should thus be included in future analyses of 
structure function data at very large~$x$.

\vspace{2mm}
\noindent
{\bf Acknowledgements:}\\[2mm]
The author gratefully acknowledges discussions with S. Catani, which 
gave rise to the research presented here. It is also a pleasure to 
thank E. Laenen and W.L. van Neerven for useful comments on the 
manuscript. This work has been supported by the European Community TMR
research network `Quantum Chromodynamics and the Deep Structure of 
Elementary Particles' under contract No.~FMRX--CT98--0194.

\newpage

\newpage
\vspace*{1cm}
\hspace*{0.5cm}
\epsfig{file=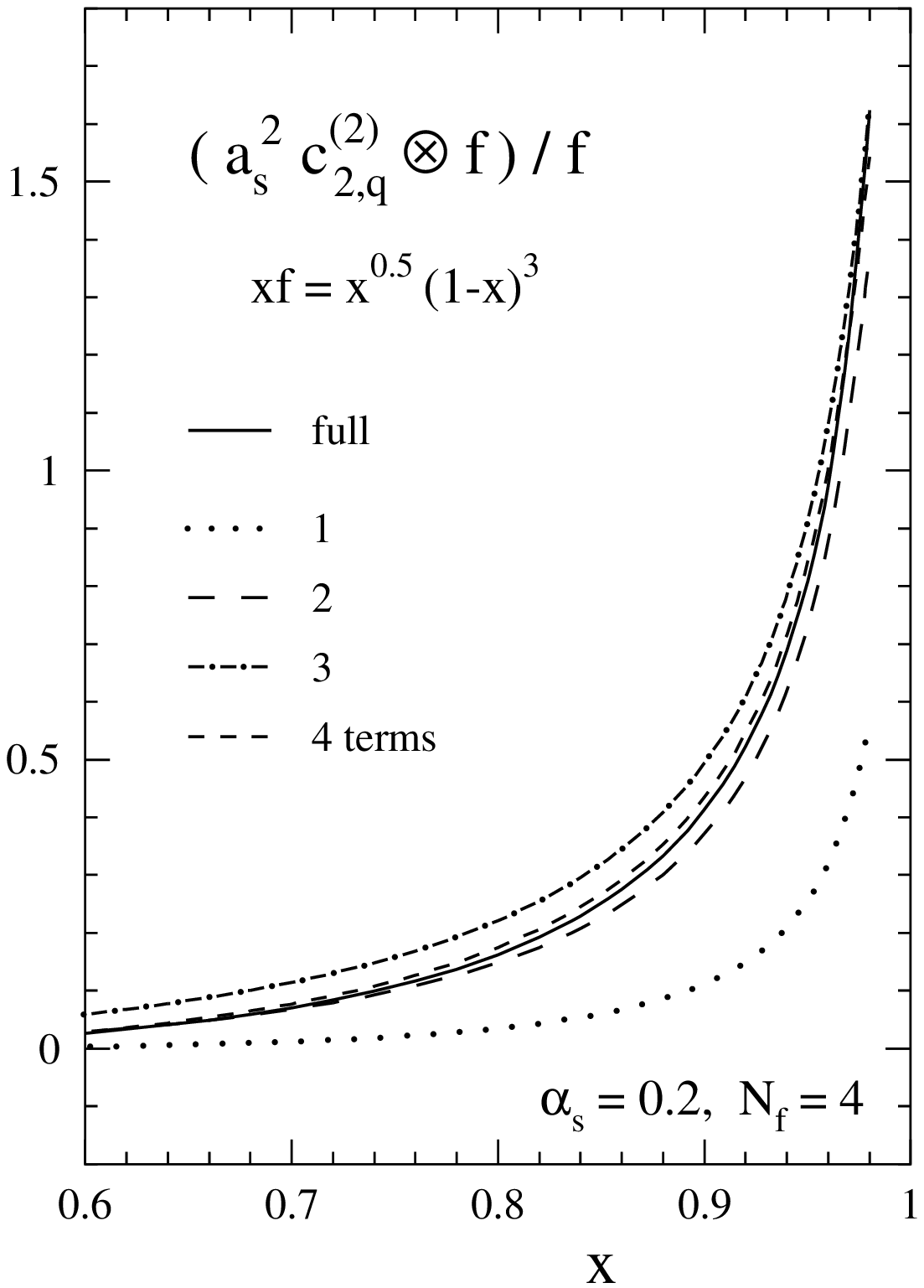,width=12cm}

\vspace{5mm}
\noindent
{\bf Figure 1:} Convolution of the second-order contribution to the 
 coefficient function $C_{2,q}$ with a typical input shape. Shown
 are the full result \cite{cq2} for $c_{2,q}^{(2)}$ and four large-$N$ 
 approximations, in which the terms $\sim \ln^4 N$, $\ln^3 N$, $\ln^2 N$
 and $\ln N$ in eq.~(\ref{coeffE}) are successively included (e.g., the 
 long-dashed curve includes the $\ln^4 N$ and $\ln^3 N$ parts).

\newpage
\vspace*{1cm}
\hspace*{0.5cm}
\epsfig{file=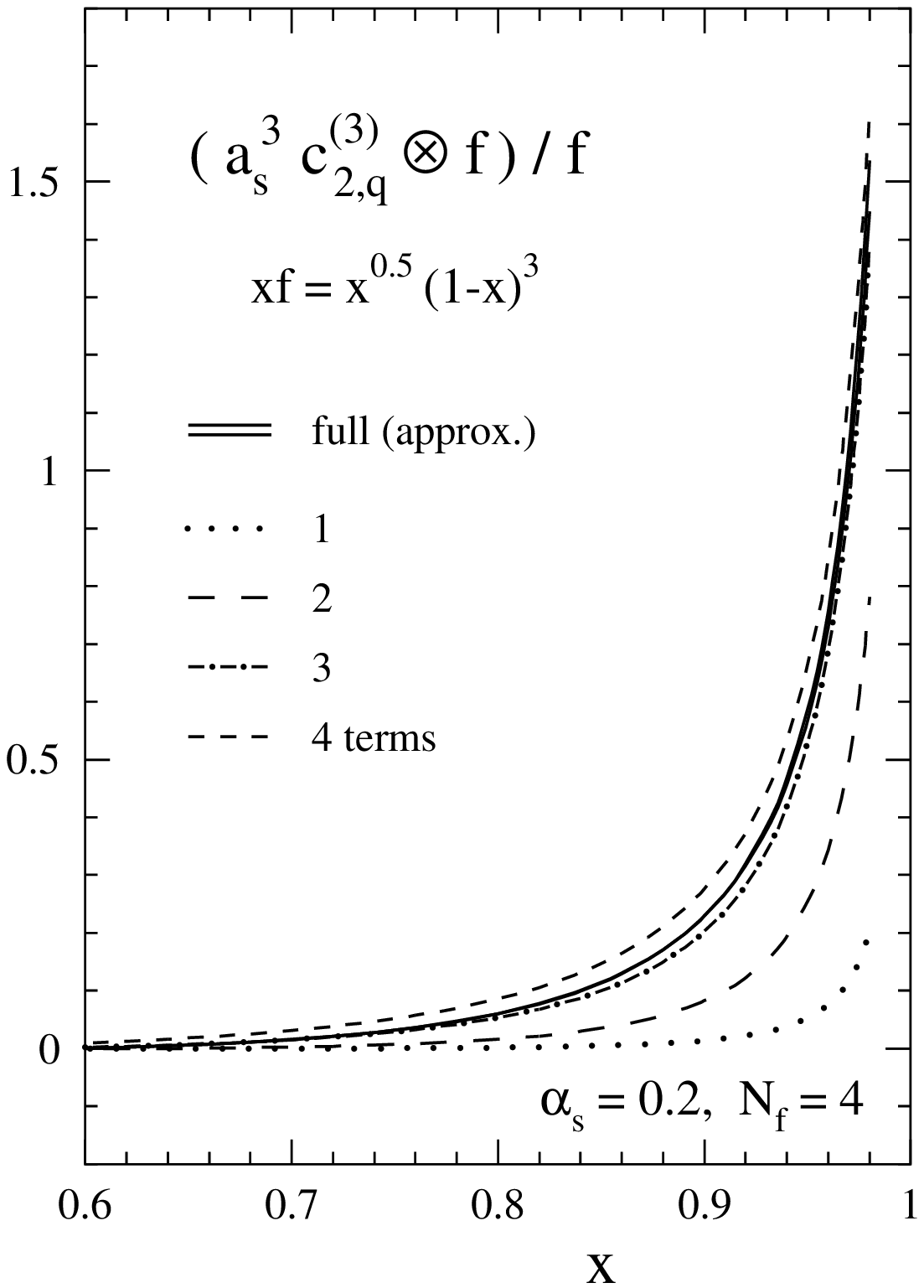,width=12cm}

\vspace{5mm}
\noindent
{\bf Figure 2:} Convolution of the third-order term $c_{2,q}^{(3)}$ of 
 the coefficient function $C_{2,q}$ with a typical input shape. The 
 four leading large-$N$ approximations are compared with a 
 parametrisation \cite{talk} of the full result based on the lowest 
 five even-integer moments \cite{LRVN}. The spread of the full curves 
 indicates the residual uncertainty of this parametrisation.

\newpage
\vspace*{1cm}
\hspace*{0.5cm}
\epsfig{file=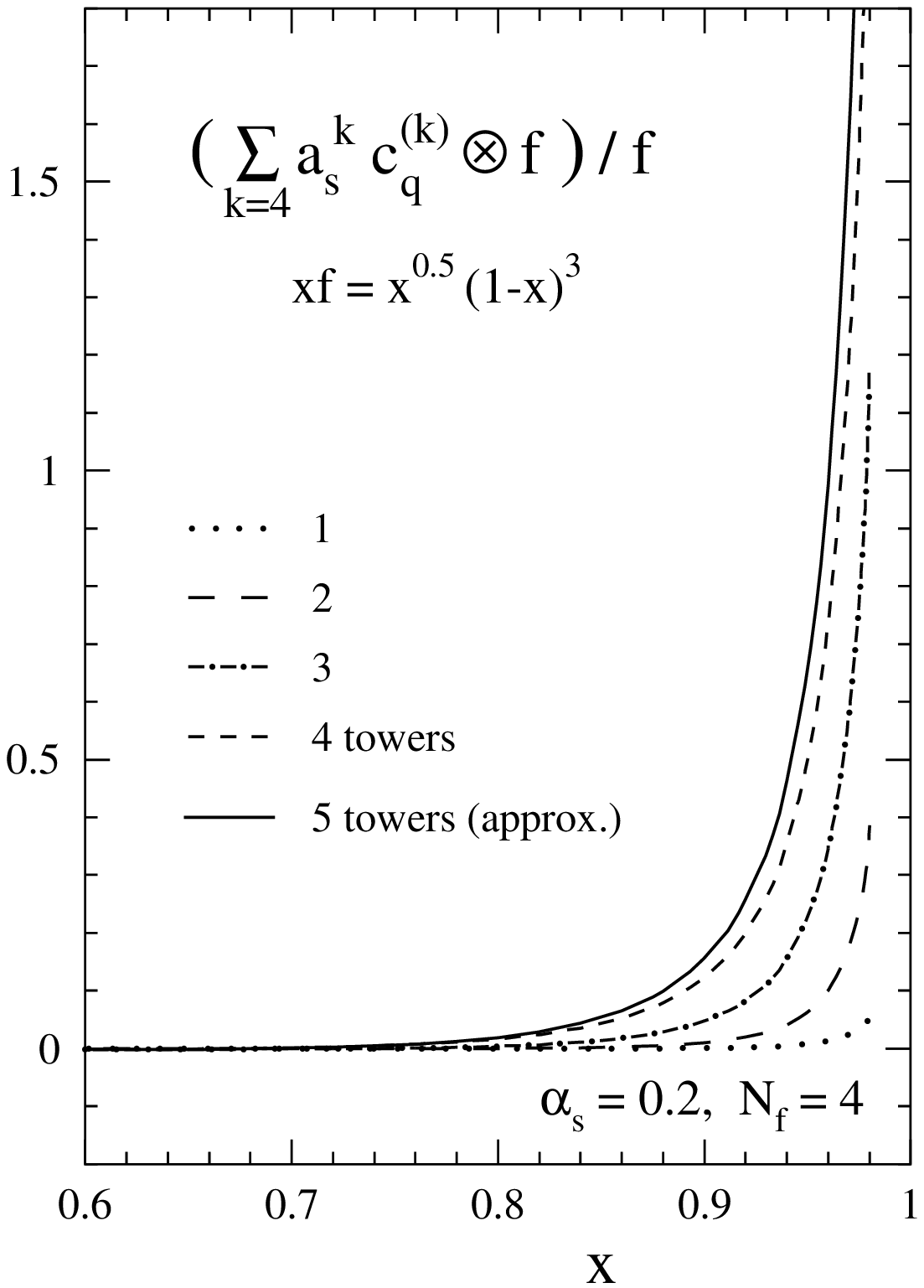,width=12cm}

\vspace{5mm}
\noindent
{\bf Figure 3:} The effect of the higher-order soft-gluon corrections, 
 $\sum_{k=4} \sum_{j=0}^{j_m} c_{kj}^{} a_s^k \ln^{2k-j} N$, to the 
 quark coefficient functions at large $N$. The results including up to 
 four towers ($j_m = 3$) are derived from the exact coefficients in 
 eq.~(\ref{towers}). The impact of the fifth tower ($j_m = 4$) is 
 estimated using $g_{33}^{} = 500$ for the coefficients $c_{k5}^{}$  
 at $k\geq 5$ in table 1.

\end{document}